\documentclass[aps,prl,
twocolumn,
superscriptaddress,
amsfonts,amssymb,amsmath,
twoside,
letterpaper,
bibnotes,
]{revtex4}

\pdfoutput=1

\usepackage{bm,color,mathrsfs,graphicx}

\newcommand{\bea}{\begin{eqnarray}}
\newcommand{\eea}{\end{eqnarray}}
\newcommand{\ben}{\begin{equation}}
\newcommand{\een}{\end{equation}}
\newcommand{\benu}{\begin{enumerate}}
\newcommand{\enu}{\end{enumerate}}

\newcommand{\om}{\omega}

\newcommand{\dl}{\delta}

\newcommand{\br}{{\bf r}}

\begin{document}

\title{Aharonov-Bohm oscillations in the local density of states}

\author{A. Cano}
\email{cano@esrf.fr}
\affiliation{
European Synchrotron Radiation Facility, 6 rue Jules Horowitz, BP 220, 38043 Grenoble, France}

\author{I. Paul}
\email{indranil.paul@grenoble.cnrs.fr}
\affiliation{
Institut N\'{e}el, CNRS/UJF, 25 avenue des Martyrs, BP 166, 38042 Grenoble, France}
\affiliation{
Institut Laue-Langevin, 6 rue Jules Horowitz, BP 156, 38042 Grenoble, France}

\date{\today}

\begin{abstract}
The scattering of electrons with inhomogeneities produces modulations in the
local density of states of a metal. We show that electron interference contributions to these modulations are affected by the magnetic field via the Aharonov-Bohm effect. This can be exploited in a simple STM setup that serves as an Aharonov-Bohm interferometer at the nanometer scale.
\end{abstract}


\maketitle

Scanning tunneling microscopy (STM) provides a powerful tool to measure the density of states \cite{Chen}. The scattering of electrons with inhomogeneities is known to modify the
corresponding density of states producing local modulations. These modulations were first
probed by STM with atomic resolution in \cite{Crommie93}, where standing-wave patterns in the local density of states (LDOS) arising from scattering of electrons with impurities and step edges were observed on the surface of Cu(111). These patterns encode information about the electronic properties of the corresponding system. For example, in the case of a two-dimensional electronic system, a point defect produces oscillations in the LDOS with wavevector $2k_F$ for very low bias volatge~\cite{Crommie93}. The period of these oscillations changes as a function of the bias voltage, from which one can infer the spectrum of the electrons. In the case of a superconductor, there can be additional features reflecting the symmetry
of the superconducting gap \cite{Byers93}. Moreover, atomic manipulation permits the engineering of surfaces, making it possible to confine electrons into the so-called quantum corrals \cite{Crommie93a}. This has attracted great attention since it permits the study of lifetime effects, Kondo physics, single-atom gating, etc. \cite{Fieter03}. Very recently, open nanostructures have been shown to be amenable for quantum holographic encoding \cite{Moon09}.
In the following we show that under suitable conditions, the LDOS exhibits oscillations
due to the magnetic field that can be interpreted as due to the Aharonov-Bohm effect. Therefore the STM setup can be designed to serve as an Aharonov-Bohm interferometer at the nanometer scale.

\begin{figure}[t]
\includegraphics[width=8cm]
{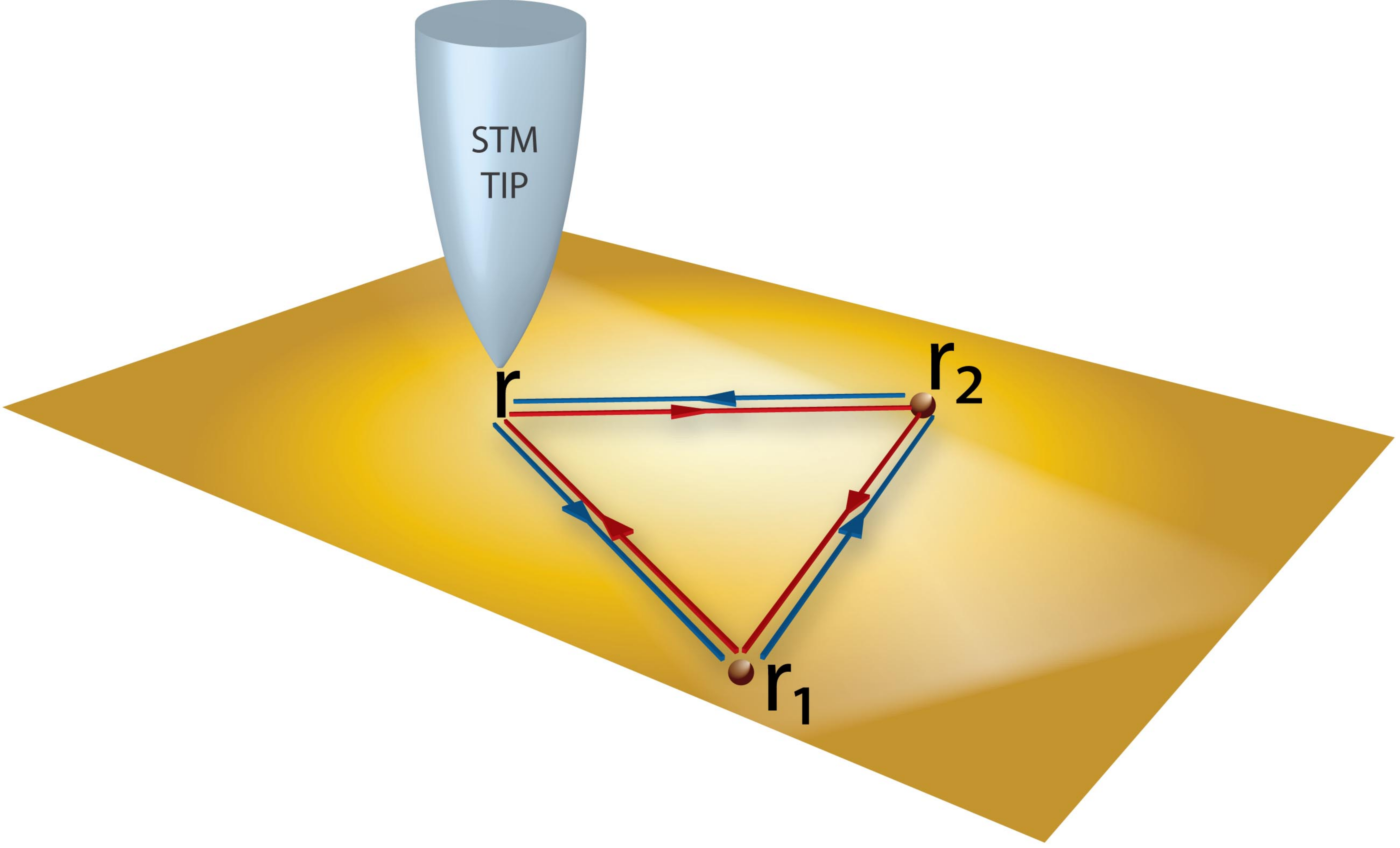}
\caption{STM interferometer. $\mathbf{r}$ represents the position of the STM tip on the surface and $\mathbf{r}_1 $ and $\mathbf{r}_2$ two impurities. The LDOS measured by the STM tip contains interference contributions due to electrons traveling along the two paths shown in the figure.
The magnetic field affects this interference via the Aharonov-Bohm effect, producing oscillations in the LDOS that is measured by the tip.}
\label{fig1}
\end{figure}

For the sake of concreteness we consider the close-packed surface of a noble metal where the so-called Shockley surface states form a two-dimensional nearly free electron gas.
Two atoms deposited on top of this surface can be modeled as two point scattering potentials for the surface electrons. This forms the simplest Aharonov-Bohm interferometer, where the role of the two different paths in conventional Aharonov-Bohm setups is played by the two scattering paths shown in Fig. \ref{fig1}. In the presence of a magnetic field, electrons scattering along these loops pick up different phases depending on whether the scattering is clockwise or anticlockwise (the two paths being connected by time-reversal symmetry at zero field).
This affects the interference contribution to the LDOS measured by the STM tip, which eventually exhibits oscillations as a function of the magnetic flux that passes through area enclosed by the above paths. The physics of these LDOS oscillations is similar to the effect of the magnetic field on weak localization in disordered two dimensional metals \cite{Altshuler80}.

The $dI(\mathbf r, V)/dV $ maps obtained experimentally are determined by the LDOS of the sample $N(\mathbf r, \omega =eV)$, the tip density of states, and the tunneling matrix elements \cite{Fieter03,Blanco06}. In the Tersoff-Hamann approximation with a constant tip density of states $dI/dV$ is proportional to $N$ \cite{Fieter03,Blanco06}. However,
the oscillations of the LDOS that we discuss are picked up by the experimental $dI /dV$
even if the above proportionality is lost due to voltage dependence of the tunneling matrix
elements or due to variation of the tip density of states with energy.
The LDOS of the sample can be obtained from the corresponding retarded Green's function as
\begin{align}
N(\mathbf r, \omega ) = -{2\over \pi}\text{Im}\, G^{R}(\mathbf r,\mathbf r; \omega)
\label{}
\end{align}
(the factor $2$ is due to spin degeneracy). For a two-dimensional free electron gas the Green's function is
\begin{align}
G^{R}(\mathbf r, \mathbf r'; \omega) = -i \pi N H_0^{(1)}\big(k({\omega})|\mathbf r -\mathbf r'|\big).
\label{G}
\end{align}
Here $N = m/(2\pi \hbar^2)$ is the density of states of the electron gas per spin,
$H_0^{(1)}$ is the zeroth-order Hankel function of the first kind and
$k({\omega})$ is given by the dispersion relation $k(\omega) = k_F (1+\omega/\mu)^{1/2}$, where $\mu=\hbar^2 k_F^2/(2m)$. For large distances [$r\gg k^{-1}({\omega})$] we have
\begin{align}
G^{R}(\mathbf r,\omega)
\approx - i N \left({2\pi\over k(\omega)r} \right)^{1/2}e^{i[k(\omega)r - \pi /4]}.
\label{G_large_r}
\end{align}

The presence of impurities can be modeled by a term
\begin{align}
H_\text{imp} = \int d\mathbf r U(\mathbf r) \rho(\mathbf r)
\label{}
\end{align}
in the Hamiltonian of the system, where $U$ is the scattering potential associated with the impurities and $\rho $ is the electronic density. Following a perturbative approach \cite{Fieter03} the Green's function can be expressed as $G = G_0 + \delta G$, where $G_0$ is the Green's function in the absence of impurities
and
\begin{widetext}\begin{align}
\delta G(\mathbf r,\mathbf r' )&=
\int d\mathbf r'' G_0 (\mathbf r - \mathbf r '') U(\mathbf r'') G_0 (\mathbf r'' - \mathbf r' )
\nonumber \\
&+ \iint d\mathbf r'' d\mathbf r''' G_0 (\mathbf r - \mathbf r '') U(\mathbf r'') G_0 (\mathbf r'' - \mathbf r''') U(\mathbf r''') G_0 (\mathbf r''' - \mathbf r' ) + \dots
\label{deltaG}
\end{align}
\end{widetext}
The dependence on the frequency is dropped since the scattering is assumed to be elastic. The latter quantity in \eqref{deltaG} contains the interference contributions to the LDOS we are interested in (i.e. from terms second order and higher in the scattering potential).

Two identical point impurities are described by the scattering potential
\begin{align}
U (\mathbf r) = U_0 [\delta (\mathbf r - \mathbf r_1) +  \delta (\mathbf r - \mathbf r_2) ].
\label{}
\end{align}
In this case the quantity $\delta G(\mathbf r, \mathbf r)$ naturally contains two types of terms. On one hand, there are
(additive) terms in which the scattering with the impurities is produced separately.
The Fourier transform of this contribution has quasiparticle peaks which contains information about the dispersion of the electrons (see e.g. \cite{Byers93}).
This contribution, however, plays no role in the physics that we intend to study.
On the other hand, there are terms involving scattering with both the impurities.
Among the latter terms, there are
processes in which the semi-classical scattering paths enclose a finite area as in Fig.~\ref{fig1}.
It is important to note that such loops occur in pairs connected by time reversal symmetry (i.e., clockwise and anti-clockwise). For example, to the lowest order in the impurity potential, the contribution to the Green's function due to closed loops reads
\begin{align}
\delta G_\text{loop}^{(2)}(\mathbf r , \mathbf r) & =
U_0^2 G_0 (\mathbf r - \mathbf r_1) G_0 (\mathbf r_1- \mathbf r_2) G_0 (\mathbf r_2 - \mathbf r )
\nonumber \\
& \quad + (1 \leftrightarrow 2).
\label{interference12}
\end{align}
In the absence of a magnetic field the two terms in this expression are equivalent (due to time reversal symmetry), and therefore give the same contribution to the LDOS (it can be said that the interference is constructive). In the presence of a magnetic field, however, time reversal symmetry is broken. Then electrons travelling clockwise and anti-clockwise along the above loop acquire different phases, so the subsequent interference is affected. This results in Aharonov-Bohm oscillations of the LDOS as we shall see explicitly below.

Next we demonstrate that the simple geometrical picture above remains unchanged when
higher order terms in the impurity potential are taken into account. At higher order
we have to deal with the following additional ingredients: (i) multiple scattering at the impurities and (ii) multiple scattering where
semi-classically the particle goes back and forth between the two impurities. The first
type of multiple scattering can be easily taken into account by replacing each of the
scattering potentials by their respective $T$-matrices. That is \cite{note},
\begin{align}
U_0 \to \widetilde U_0 = \frac{U_0}{1 - U_0 G_0 (0) }.
\label{U_0}\end{align}
As regards the second point, we note that $\dl G_{\rm loop}(\br, \br)$ is entirely due to processes where the path between the impurities is traversed an odd number of times (otherwise the scattering path does not enclose a finite area).
In the second order contribution \eqref{interference12}, for example, the path between the impurities is traversed once. At ${\mathcal O}(U_0^4)$ there are contributions where the path is traversed three times, and so on. Taking all this into account, we obtain
\begin{align}
\dl G_{\rm loop}(\br, \br)
&= W^2
G_0 (\br - \br_1) G_0 (\br_1 - \br_2) G_0 (\br_2 - \br)
\nonumber \\
&\quad + (1 \leftrightarrow 2),
\label{Gloop}
\end{align}
where
\begin{align}
\label{W}
W^2 = \frac{\widetilde U_0^2}{1 - \widetilde U_0^2 G_0(\br_1-\br_2) G_0(\br_2-\br_1)}.
\end{align}
It is worth noticing that in all the processes that finally give rise to \eqref{Gloop} we are actually dealing with the same area, since to go back and forth along the same line, for example, does not change the area of the resulting loop. This is the simple reason why the non-trivial phase relation between the clockwise and the anticlockwise paths in the
presence of a magnetic field is not washed out by multiple impurity scattering.

Let us now consider explicitly the influence of the magnetic field. In the low-field regime (see below) we can use the semi-classical approximation for the electron Green's function
\cite{Altshuler80}:
\begin{align}
G_0 (\mathbf r - \mathbf r') = \exp\left({i{\pi \over \Phi_0}
\int_{\mathbf r}^{\mathbf r '}\mathbf A (\mathbf l)\cdot d \mathbf l }\right) G_{00} (\mathbf r- \mathbf r').
\label{geometricalG}
\end{align}
Here $G_{00} (\mathbf r- \mathbf r')$ represents the Green's function in the absence of magnetic field, $\mathbf A $ is the vector potential ($\mathbf B =\nabla \times \mathbf A$), $\Phi_0 = h/(2e)$ is the flux quantum, and the integral is along the straight line connecting $\mathbf r$ and $\mathbf r'$. The magnetic field then enters
the interference contributions to the LDOS via complex factors $e^{\pm i\pi \Phi/\Phi_0}$, where $\Phi$ is the
magnetic flux through the area enclosed by the corresponding scattering path (the different signs of the phase corresponds
to anti-clockwise and clockwise line integrals respectively). As a result, the LDOS can be written as
\begin{align}
N(\mathbf r,\omega) = N_{B =0}(\mathbf r,\omega) + N_\text{loop}(\mathbf r,\omega)
[\cos(\pi \Phi/\Phi_0) - 1],
\label{LDOS}
\end{align}
where $N_{B =0}$ is the zero-field total LDOS and
\begin{align}
N_{\rm loop}(\br, \om) =
- \frac{2}{\pi} {\rm Im} \, \dl G_{\rm loop}(\br, \br;\omega)
\end{align}
represents the (constructive) interference contribution due to all closed paths in the absence of a magnetic field [computed from \eqref{Gloop} with $G_{00}$]. This interference process picks up a non-trivial phase in the presence of a magnetic field, giving rise to oscillations in the LDOS as in \eqref{LDOS},
which is a manifestation of the Aharonov-Bohm phenomenon \cite{note-TH}. This can be revealed by varying either the magnetic field or the relative position between the STM tip and the impurities since $\Phi$ changes in both cases. We note that, irrespective of the position of the STM tip, the magnitude of the correction to the LDOS reduces
with $B$ for low magnetic fields. As in the case of negative magnetoresistance in weak localization, this is due to the fact that magnetic field induces destructive interference between the contribution of the two semi-classical paths.

Next we discuss the limitations of our calculations and the feasibility of our proposal to use the STM as an Aharonov-Bohm interferometer at a nanometer scale. The expression \eqref{LDOS} for the LDOS has been derived within a semi-classical approach, so it holds as long as Eq. \eqref{geometricalG} can be used for describing the influence of the magnetic field on the electron system. This is possible if the magnetic field is such that the Fermi
wavelength is much smaller than the Landau orbits:
\begin{align}
\lambda_F \ll a_B =
\left({\Phi_0\over \pi  B }\right)^{1/2}.
\label{semiclassic}
\end{align}
The magnetic field needed to observe a complete Aharonov-Bohm oscillation can be estimated as
\begin{align}
B_{2 \pi} \sim 4 \Phi_0/d^2,
\end{align}
where $d$ is the characteristic distance in the setup (i.e. the distance between the impurities
and/or between the impurities and the position of the tip. For $d \sim 40-20 \, \rm nm$ this field is $B_{2 \pi} \sim 5-20\, \rm T$,
which corresponds to Landau orbits $a_B \sim 11-6 \, \rm nm$.
For Cu(111) and Ag(111) $\lambda_F = 2.95\,\rm nm$ and $7.6\,\rm nm$ respectively, and therefore the condition in Eq.~\eqref{semiclassic} for the semi-classical approximation is valid. Furthermore, inspite of the fact that the interference signal is long-range in the sense that $N_\text{loop}$ decays as $1/d$ in a power-law fashion [see \eqref{G_large_r}], in reality there are dephasing processes that introduce an extra attenuation (thermal dephasing for example), and which, for simplicity, have not been taken into account in the current computation. Experimentally, in fact, the impurity-induced variations in the LDOS are typically observed up to distances of the order of a few times the Fermi wavelength, say $10 \lambda_F$ \cite{note-dephasing}. Therefore the characteristic distance in our setup must be $d \lesssim 10 \lambda_F$. For $d \sim 40-20 \rm nm$ we then need $\lambda_F \gtrsim 4-2 \rm nm$, which still can be smaller than $a_B$ at $5-20\, \rm T$.
Thus we find that the semiclassical interpretation is good to describe the first few periods of the LDOS oscillations. 
The first corrections to our results will be due to the curvature of the classical trajectories, which can still be described within the semiclassical picture \cite{Sedrakyan07}.
To describe sufficiently high periods, however, one has to go beyond \eqref{geometricalG} and consider the influence of the magnetic field within a Landau-level approach. These periods imply magnetic fields considerably high ($>20 \, \rm T$), so we do not develop this latter approach here.

\begin{figure}[t]
\includegraphics[width=.235\textwidth]
{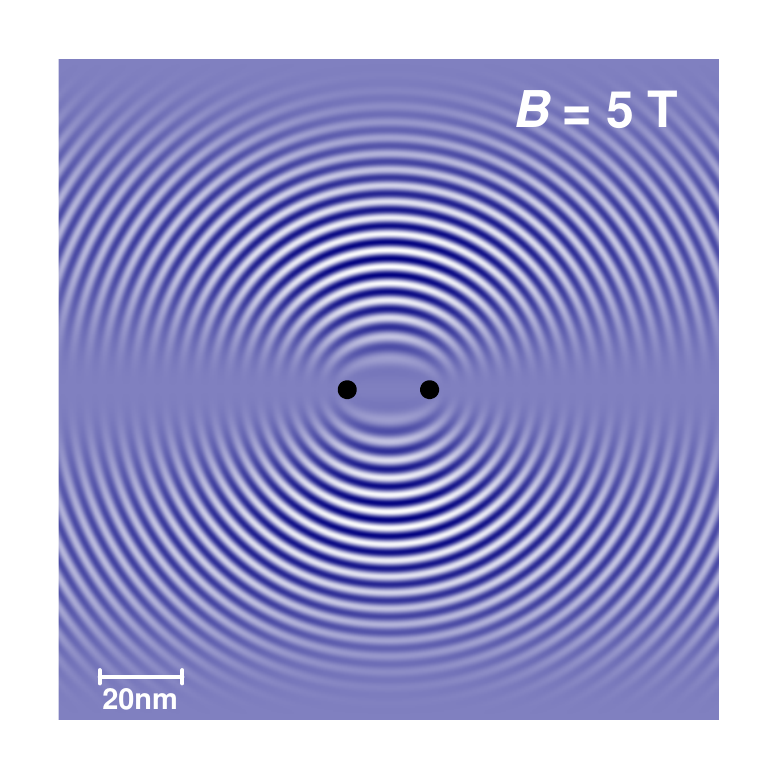}
\hspace{.005\textwidth}
\includegraphics[width=.235\textwidth]
{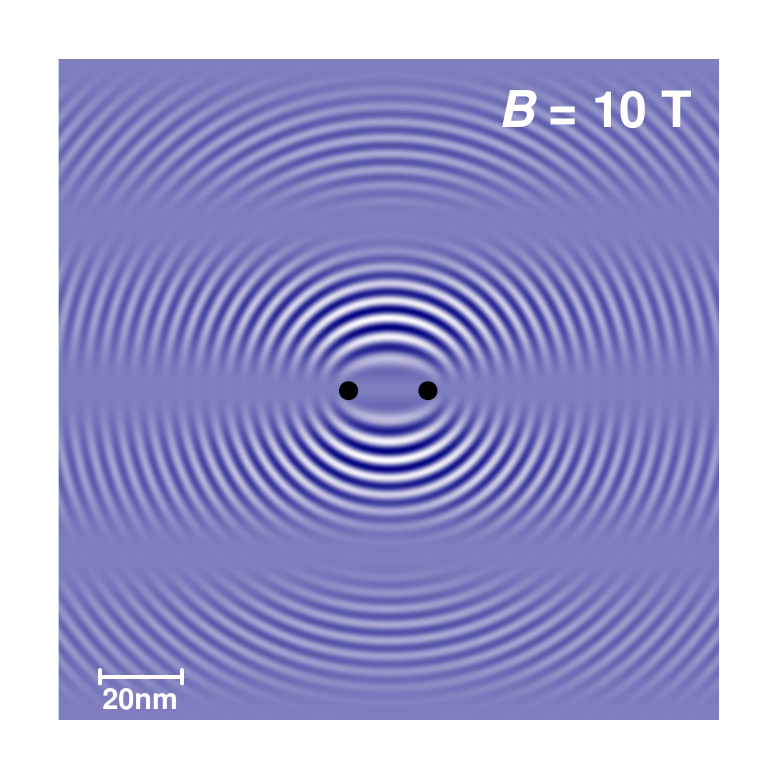}
\hspace{.005\textwidth}
\includegraphics[width=.235\textwidth]
{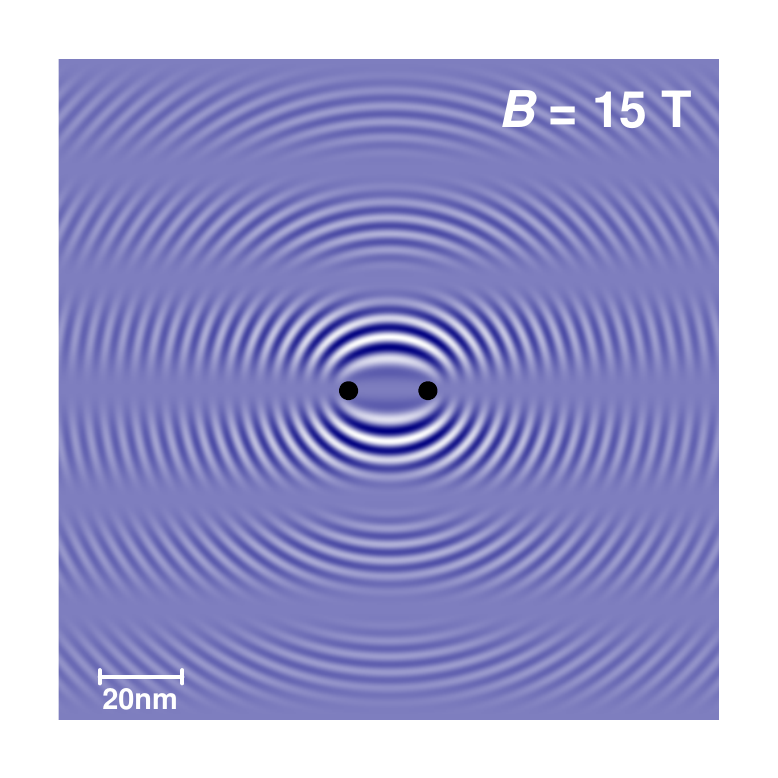}
\hspace{.005\textwidth}
\includegraphics[width=.235\textwidth]
{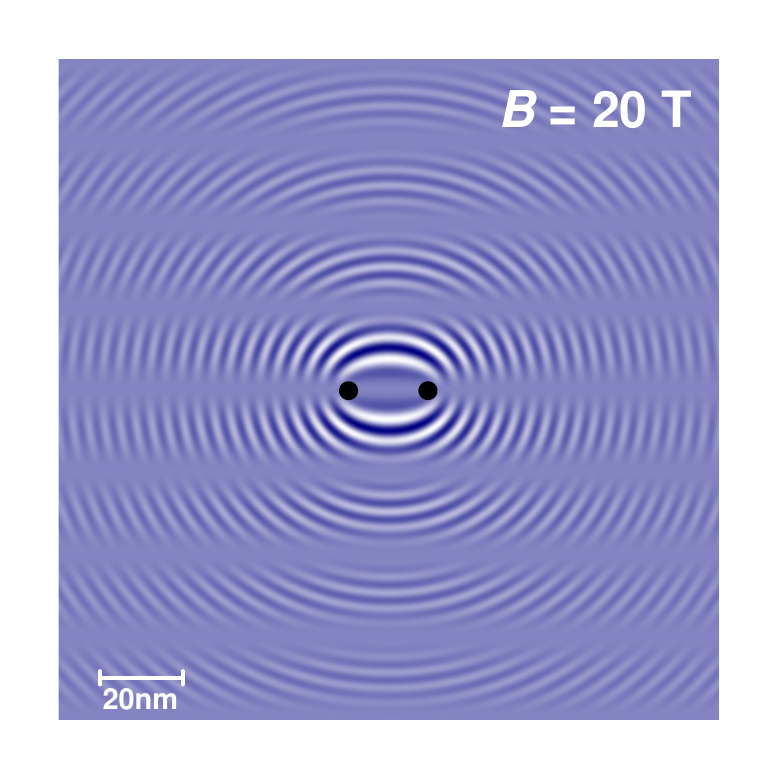}
\caption{Expected STM patterns for two impurities $20 \, \rm nm$ apart on the Ag(111) surface after subtraction of the $B=0$ signal.
The horizontal stripes are produced by the Aharonov-Bohm effect in the LDOS, whereas the remaining elliptic features are due to the interference contributions described by $N_\text{loop}$ [with $\lambda_F = 7.6\, \rm nm$ and real $W$-matrices as defined in Eq.~\eqref{W}].}
\label{STMmaps}
\end{figure}

The Aharonov-Bohm physics reveals also in the spatial variations of the LDOS for a fixed magnetic field. In our setup $\Phi$ varies only in the direction perpendicular to the line connecting the two impurities. In consequence, the STM scans along this direction will show a periodic envelope due to the cosine factor in Eq. \eqref{LDOS}. With two impurities separated $20 \, \rm nm$, fields of $5 - 20 \, \rm T$ give rise to periods of $80$ and $ 20\, \rm nm$ respectively for such an envelope (see Fig. \ref{STMmaps}).

It is also worth mentioning that the Zeeman splitting, neglected so far, has a trivial influence on the Aharonov-Bohm oscillations if the spin is conserved in the scattering process. $N_\text{loop}$ in Eq. \eqref{LDOS} actually results from contributions associated with the two spin polarizations. It can be written as $N_\text{loop}={1\over 2}(N_\text{loop}^\uparrow + N_\text{loop}^\downarrow)$ if the spin is conserved. The eventual difference between $N_\text{loop}^\uparrow $ and $N_\text{loop}^\downarrow $ due to the Zeeman splitting can be probed by means of spin polarized STM. However this difference does not alter the oscillatory behavior of the LDOS described above. The situation is more subtle if the spin can flip during the scattering process as a result of e.g. spin-orbit coupling. This can affect multiple scattering in a nontrivial way \cite{Walls07}, and can give rise to the analog of the anti-localization phenomenon if spin components are measured separately.

In summary, we have shown that magnetic field affects electron interference contributions to the local density of states via the Aharonov-Bohm effect. This can be exploited in building STM devices that serve as Aharonov-Bohm interferometers at the nanometer scale. We have illustrated this possibility for the close-packed surface of a noble metal with two atoms adsorbed on top. The role of these atoms consists in creating strong enough scattering potentials, which can also be produced, for example, using additional STM tips. The implementation of this new functionality into the STM technique might broaden its applications notably, offering new perspectives for STM studies of the fundamental properties of surfaces and underlying systems.

We acknowledge I. Brihuega, P. Bruno, E. Kats and R. Whitney for very fruitful discussions and M. Collignon for the Figure 1.

\onecolumngrid

\twocolumngrid


\begin{references}
\bibitem{note-TH} Note that the experimental $dI/dV$ maps will show the Aharonov-Bohm oscillations even if $dI/dV$ is not proportional to the local density of states of the sample.
\end{references}
\end{document}